\documentclass[12pt]{iopart}

\usepackage{graphicx}

\usepackage{iopams}

\newcommand{\eqref}[1]{(\ref{#1})}
\newcommand{\bea}{\begin{eqnarray}}
\newcommand{\eea}{\end{eqnarray}}
\newcommand{\beq}{\begin{equation}}
\newcommand{\eeq}{\end{equation}}

\newcommand{\bal}{\begin{align}}
\newcommand{\eal}{\end{align}}
\newcommand{\bit}{\begin{itemize}}
\newcommand{\eit}{\end{itemize}}

\newcommand{\cD}{{\cal D}}

\def\cmp{{\complement}}

\def\tA{{\tt A}}

\def\tz{{\tt z}}

\def\Om{\Omega}

\def\bom{{\mbox{\boldmath$\omega$}}}
\def\obom{\overline\bom}
\def\0bom{{\bom}^0}
\def\0obom{{\obom}^0}

\def\0nbom{{\bom}_{n,0}}
\def\n*bom{{\bom}^*_{(n)}}

\def\Lam{\Lambda}

\def\bbP{\mathbb P}

\def\bx{\mathbf x}
\def\ubx{\underline\bx}

\def\by{\mathbf y}

\def\uby{\underline\by}

\def\cD{\mathcal D}

\def\cS{\mathcal S}

\def\bbR{\mathbb R}

\def\bbZ{\mathbb Z}

\def\bx{\mathbf x}

\def\ua{{\underline a}}

\def\oa{{\overline a}}

\def\by{\mathbf y}

\def\unn{\underline n}

\def\bx{\mathbf x}

\def\diy{\displaystyle}

\def\u0{{\underline 0}}

\def\rd{{\rm d}}

\def\rd{{\rm d}}

\def\oa{\overline a}
\def\ua{\underline a}

\newtheorem{theorem}{Theorem}
\newtheorem{proposition}[theorem]{Proposition}

\begin{document}

\title[Dominance of most tolerant species in multi-type lattice WR models]{Dominance of most tolerant species in multi-type lattice Widom-Rowlinson models}

\author{A.~Mazel$^1$, Yu.~Suhov$^2$, I.~Stuhl$^3$, S.~Zohren$^4$}

\address{
$^1$ AMC Health, New York, NY, USA \\
$^2$ Statslab, DPMMS, University of Cambridge, UK, \\
$^3$ University of Debrecen, Hungary, \\
$^4$ Department of Physics, Pontifica Universidade Cat\'olica, Rio de Janeiro, Brazil
}



\begin{abstract}
We analyse equilibrium phases in a multi-type lattice Widom--Rowlinson model with (i) four particle types,
(ii) varying exclusion diameters between different particle types and (iii) large values of fugacity. Contrary 
to an expectation, it is not the most ``aggressive'' species, with largest diameters, which dominates the 
equilibrium measure, but  the ``most tolerant'' one, which has  smallest exclusion diameters. Results of 
numerical simulations are presented, showing densities of species in equilibrium phases and confirming
the theoretical picture.
\end{abstract}

\maketitle

\section{Introduction}
%
The \emph{Widom--Rowlinson} (WR) model was introduced in 1970 \cite{WR} as a simple continuum statistical physics 
model to study the thermodynamic equilibrium of a classical fluid in which one can rigorously analyse the existence of a liquid-vapor phase transition \cite{Ru}. In its original formulation the WR-model is given in terms of a single type of particles in a continuum $d$-dimensional space which have a specific interaction potential. It can be mapped onto a two-type mixture where particles of the same type do not interact and particles of different type interact according to a given pairwise hard-core exclusion diameter (ED) $D$. Subsequently a lattice version of the model was introduced \cite{LG} as well as a generalisation for multiple types  \cite{RL}. Aspects of these models have 
been studied e.g.\ in 
\cite{LG,RL,CCK,LMNS,DS,NL,GZ}
(see also the bibliography therein). In the case of a multi-type model it was assumed that different particles have the same EDs. Recently \cite{MSS} a variation of the multi-type continuum WR-model was introduced, with varying EDs between different particle types. Such models are interesting mathematically and may have applications in other fields including biology, medicine and sociology.

Biological examples of a multi-type WR-model are organism exhibiting \emph{allelopathy} \cite{ID} such as certain algae, bacteria, coral, and fungi.  A similar pattern is also related to growth of cancer cells lowering usual levels of tolerance \cite{ZLXHH,App1}. A natural question is which species will dominate after a long growth time when we reach an equilibrium measure, under a high rate of growth. A na\"ive guess might be that the most ``aggressive'' species, with largest EDs, will be predominant. But 
in fact,  \emph{the equilibrium measure is dominated by a ``most tolerant'' species}, with  smallest EDs. A detailed understanding of this phenomenon is the main motivation of this work.

We analyze a lattice version of the $q$-type WR-model in $\bbZ^d$, 
for a large (and symmetric) fugacity $\tz$, in analogy with a continuum model \cite{MSS}, adopting its notational system and terminology. Following \cite{MSS}, for $q=2,3,4$ we give a description of equilibrium 
measures in Proposition 1 (i) below. In addition, we analyze in detail the particle densities in these measures, cf.\ Proposition 1 (ii). Working in a lattice 
has an advantage of complementing theoretical results with numerical studies by using a suitable spin-flip process. 

\section{The $q$-type lattice WR-model}
%
We consider particles placed at nodes of a lattice $\bbZ^d$, with
the maximal occupancy number per site $1$, i.e.,\ at each site there is one or no particle. (There 
is a version of the model where the number of particles at a single node is unbounded; our theoretical 
results are extended to this case without major modifications.)
In addition, each particle has an attached type $i=1$,..., $q$, and no two particles of types $i$,
$j$ with $i\neq j$ can occupy sites $y,y^\prime\in\bbZ^d$ with ${\rm{dist}}\,(y,y^\prime)\leq D(i,j)=D(j,i)$.
Here $\mathrm{dist}(\cdot,\cdot)$ stands for the graph distance in $\bbZ^d$, and $\cD =\{D(i,j)\}$ is a given 
collection of positive integers representing hard-core EDs. 

An individual configuration 
is identified with a function $\ubx :y\in\bbZ^d\mapsto \ubx (y)\in\{0,1,... ,q\}$.
The configuration space of the model is denoted by $\tA$; formally,
\bea\label{eq:1.1}
\begin{array}{r}\tA=\Big\{\ubx:\;{\rm{dist}}\,(y,y^\prime) > D(i,j)\;\forall\;y,y^\prime\in\bbZ^d
\qquad\qquad{}\\
\hbox{with}\;\ubx (y)=i,\ubx (y^\prime )=j,\;\;
\forall\;1\leq i<j\leq q\Big\}.\end{array}
\eea

Set $\Om =\{0,1,\ldots q\}^{\bbZ^d}$ and let $\bbP$ denote a Bernoulli distribution where each 
site $y\in\bbZ^d$ gets a label $0$ with probability $1-p$ and a label $j\in\{1,..., q\}$ with probability 
$p/q$ where $p\in (0,1)$ is a given number. Let $\Lam =\Lam_L$ denote a lattice cube: 
$\Lam =[-L,L]^d\cap\bbZ^d$ where $L$ is a positive integer. Given $\uby\in\tA$, define: 
\bea
\label{eq:1.2}
\tA_\Lam (\uby)=\{\ubx\in\Om :\;\ubx^\Lam\vee\uby^{\cmp\Lam}\in\tA\}.
\eea
Here $\ubx^\Lam$ denotes the restriction of $\ubx$ to $\Lam$, $\uby^{\cmp\Lam}$
the restriction of $\uby$ to the complement $\cmp\Lam =\bbZ^d\setminus\Lam$ and 
$\ubx^\Lam\vee\uby^{\cmp\Lam}$ stands for the concatenation of the
two restrictions. Then $\bbP (\tA_\Lam (\uby))>0$, and we consider the conditional distribution
\beq\label{eq:1.3}
\mu_\Lam (\,\cdot\,||\uby )=\frac{\bbP (\,\cdot\,\cap\tA_\Lam (\uby))}{\bbP (\tA_\Lam (\uby))}.
\eeq
In the literature, $\mu_\Lam (\,\cdot\,||\uby )$ is called the \emph{Gibbs distribution} in $\Lam$ 
with a \emph{boundary condition} $\uby$. This distribution is associated to the 
\emph{grand canonical ensemble}. When the configuration $\uby$ has $\uby (x)\equiv i$ 
$\forall$ $x\in\bbZ^d$, we write $\mu_\Lam (\,\cdot\,||i)$
instead of $\mu_\Lam (\,\cdot\,||\uby)$.

We study \emph{DLR} (Dobrushin-Lanford-Ruelle) \emph{measures} $\mu$ \cite{Dobrushin,LR} 
supported by the event $\tA$ and satisfying the DLR equation (where $\mu$ is an unknown):
\beq\label{eq:1.4}
\int f(\ubx^\Lam) \mu (\rd\ubx)=\int\mu(\rd \uby)\int f(\ubx^\Lam)\mu_\Lam(\rd\ubx||\uby).
\eeq
Here $\Lam$ is a cube and $f:\tA\to [0,+\infty )$ a function depending upon the 
restriction $\ubx^\Lam$. The DLR measures are obtained from the distributions $\mu_\Lam (\,\cdot\,||\uby)$ via the \emph{thermodynamic limit}, as $\Lam\nearrow\bbZ^d$.

In our approach we keep fixed the value $q$ and the collection $\cD$ 
(assumed to be of a general form) 
and let $p$ be close to $1$. Thus, the fugacity $\tz=p/(1-p)$ is large. Moreover, as in 
\cite{MSS}, our results are given for $q\leq 4$;  for 
definiteness we focus on $q=4$. A major issue are 
\emph{pure phases} (PPs), i.e., shift-periodic ergodic DLR measures emerging as limits of distributions
$\mu_\Lam (\,\cdot\,||i)$ for \emph{dominant} types $i$. (The so-called staggering phases 
will not occur in our setting, cf.\ \cite{GZ} and references therein.) It is well-known
that for $p\sim 0$ one has a unique DLR measure $\mu$. This measure describes an infinite connected
external cluster (a ``sea'') of empty sites
with (relatively rare) ``islands'' of occupied sites, inside which there may be ``lakes'' of empty sites
containing smaller islands of occupied sites, etc. On the other hand, for $p\sim 1$ both uniqueness and
non-uniqueness of a PP can occur but the picture differs: the external cluster is occupied by a 
dominant particle type.  A passage from one pattern to another, for a given pair $(q,\cD)$ but 
varying $p$, is interpreted as a \emph{phase transition}.

\begin{figure}[t]
\centering 
\includegraphics[width=4cm]{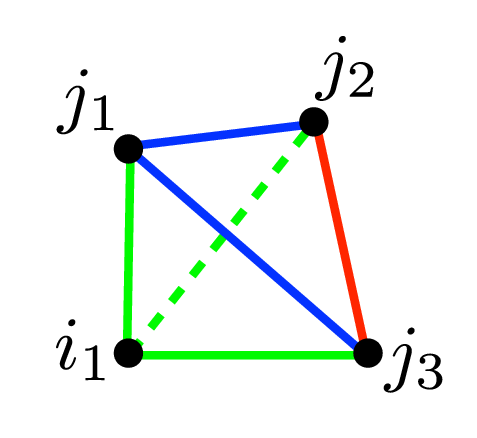}
\caption{Pure phases on a colored tetrahedron in $\bbR^3$. Here, for $p\sim 1$, a single PP occurs, and species $i_1$ dominates the (unique) equilibrium measure. Indeed, $i_1$ is the most tolerant species, as indicated by the smallest  EDs, shown in green.}
\label{fig1}

\end{figure}

\section{Dominant species in equilibrium phases}
%
For $q\leq 4$, as in \cite{MSS}, we list the hard-core EDs in 
increasing order, omitting repetitions: $\ua =a(1)<a(2)<\ldots <a(k)=\oa$.
Here  $k=1,...,6$ is the number of pair-wise distinct values among EDs $D(i,j)$. Define the 
vector $\unn(i)$ with entries $n_l(i)$, $1\leq l\leq k$:
\bea\label{eq:2.2}
n_l(i)=\sharp\,\Big\{j\in\{1,2,3,4\}\setminus\{i\}: D(i,j)=a(l)\}\Big\},
\eea
giving the occupation numbers of the EDs for type $i$. Take 
the collection $\cS^{} \!\equiv\! \cS^{}(\cD )$ of types $i$ for which the vector $\unn (i)$ is 
lexicographically maximal (beginning with $n_1(i)\geq n_1(j)$):
\beq\label{eq:2.3}
\cS^{} =\left\{i:\;1\leq i\leq 4,\;\unn (j)\buildrel{\rm{lex}}\over \preceq\unn (i)
\;\forall\;1\leq j\leq 4\right\}.
\eeq

\begin{proposition}
As in {\rm\cite{MSS}}, assume the triangular condition
\beq\label{eq:1.0}
D(i,k)\leq D(i,j)+D(j,k),\;1\leq i,j,k\leq 4.
\eeq
There exists a $p^*\in (0,1)$ such that for all $p\in (p^*,1)$,
we have the following properties:

(i)  For $i\in\cS$, there exists the limiting measure
\beq\label{eq:3.A}
\mu (\,\cdot\,||\,i)=\lim\limits_{\Lam\nearrow\bbZ^d}\mu_\Lam(\,\cdot\,||\,i).
\eeq
The measures $\mu (\,\cdot\,||\,i)$ are pairwise disjoint shift-invariant PPs. Any
shift-periodic DLR measure $\mu$ is a mixture of  $\mu (\,\cdot\,||\,i)$, $i\in\cS$. In
particular, for all 
$j\in\{0,1,2,3,4\}\setminus\cS$, the distributions $\mu_\Lam (\,\cdot\,||\,j)$ converge to a
mixture of measures $\mu (\,\cdot\,||\,i)$, with coefficients determined as in \cite{MSS}. 
The measure $\mu(\,\cdot\,||i)$ has a unique infinite connected cluster of sites
occupied by particles of type $i$; other species and empty sites produce finite clusters.

(ii) Given an $i\in\cS$ and $0\leq j\leq 4$, a cube $\Lam$ and a site 
$y\in\Lam$, set $\rho_\Lam(y,j\|i)=\mu_\Lam(\ubx (y)\!\!=\!\! j||\,i)$.
The following limits hold:
\beq\label{eq:3.2}
\rho (j\|i)=\mu\big(\ubx (y)\!=\! j||\,i\big)=\lim_{\Lam\nearrow\bbZ^d}\rho_\Lam(y,j\|i).
\eeq
The value $\rho (j\,||\,i)$ represents the type $j$ particle density in the measure 
$\mu (\;\cdot\;||\,i)$ whereas $\rho (0\,||\,i)$ yields the ``vacuum'' density. The 
values $\rho (j\,||\,i)$
satisfy a system of relations depending on the collection of the EDs $\cD$. Viz., for
$i',i''\in\cS\setminus\{i\}$, $j\not\in\cS$,
\beq\label{eq:3.2A}
\rho (i\|i)>\rho (i'\|i)=\rho (i''\|i)>\rho (j\|i).
\eeq
As $p\nearrow 1$, monotonically with $p$,
\beq\label{eq:3.3}
\rho (i\,||\,i)\nearrow1, \quad \rho (j\,||\,i)\searrow 0,\;j\neq i.
\eeq
\end{proposition}

A detailed mathematical proof of Proposition 1 will be given elsewhere. 
Instead, in the next section, we provide numerical evidence for the above proposition.

To understand 
this result,  we visualise the particle types as vertices of
a tetrahedron in $\bbR^3$; cf. Figure \ref{fig1}. The edges 
are associated with $D(i,j)$. (We do not mean 
the length of an edge equals $D(i,j)$.)
The number $k\leq 6$ of distinct values for the EDs gives 
the number
of different ``colors'' used for painting the edges. For instance, $a(1)$ can be
painted green, $a(2)$ blue, $a(3)$ red and so on. 

In fact, homeomorphic colored pictures lead to equivalent models, with
the same set of PPs. The total number
of non-homeomorphic pictures equals $299$: $60$ six-colored,
$120$ five-colored, $74$ four-colored, $36$ three-colored,  $8$
two-colored, $1$ single-colored. 

\begin{figure}[t]
\centering 
\includegraphics[width=8cm]{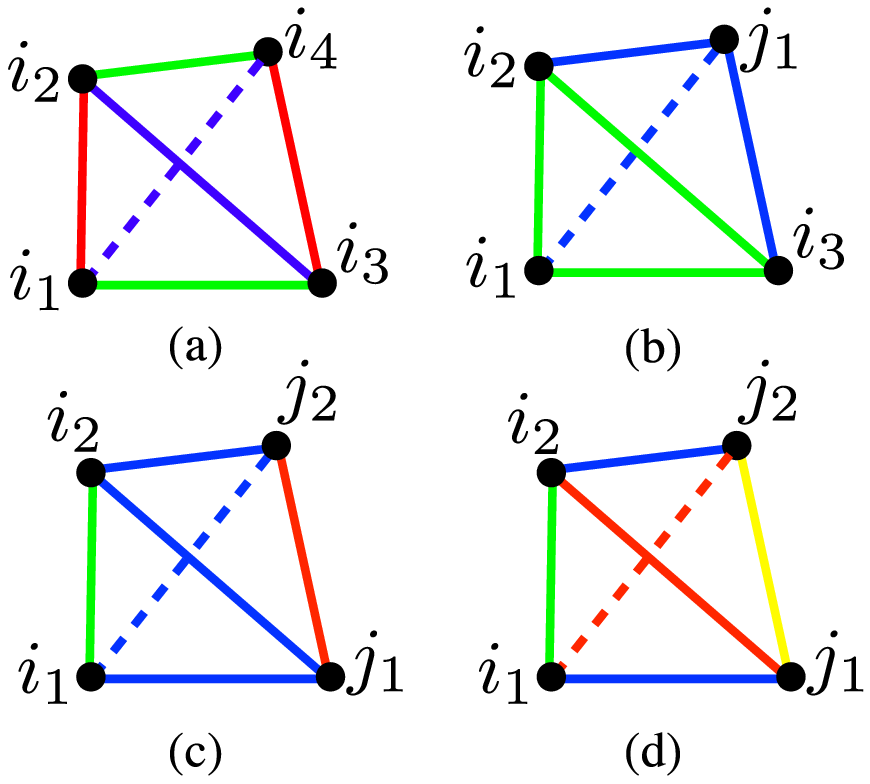}
\caption{Examples of configurations with more than one PP: (a) four PPs, (b) 
three PPs, (c) and (d) two PPs.}
\label{fig2}
\end{figure}

The number of different PPs for a given $\cD$ is as follows.
\begin{enumerate}

\item 
Four PPs
occur in 4 pictures: 1 single-color, 2 two-color, 1 three-color (as in Figure \ref{fig2} (a)).

\item 
Three PPs
occur in 1 two-color picture (as in Figure \ref{fig2} (b)). %

\item 
Two PPs occur in
27 pictures: 2 two-color, 10 three-color  (see Figure \ref{fig2} (c) for an example), 
15 four-color  (see Figure \ref{fig2} (d) for an example).%

\item 
Finally, a single PP occurs in the remaining
267 pictures: 3 two-color, 25 three-color  (see Figure \ref{fig1} for an example), 
59 four-color, 120 five-color, 60 six-color. 
\end{enumerate} 

For example, in Figure \ref{fig2}(c) we have PPs for the dominant species $i_1,i_2\in\cS$.
The types $j_1,j_2\not\in\cS$ are more aggressive species.  

The system of relations upon the densities $\rho (j\|i)$, including  \eqref{eq:3.2A}, is 
established via symmetry considerations.  For instance, in Figure 2(c), $\cS=\{i_1,i_2\}$, 
and for $\rho (j\|i_1)$ we have the inequalities
\beq\label{eq:4.11}
\rho (i_1\|i_1)>\rho (i_2\|i_1)>\rho (j_1\|i_1)=\rho (j_2\|i_1).
\eeq

\section{A spin-flip process for simulating the lattice WR-model}
%
Here we present numerical results obtained by using a discrete-time Markov spin-flip process. As was said, 
it can be interpreted as a physical process, describing the growth phenomenon behind the model. 

The statistical weight under distribution $\mu_\Lam (\;\cdot\;||i)$ is
\beq\label{eq:5.1}
(1-p)^{N^0_\Lam}\,(p/q)^{N^{\neq 0}_\Lam}\,{\mathbf 1}(\ubx^\Lam\in\tA_\Lam(i)), 
\eeq
where $N^0_\Lam=N^0_\Lam (\ubx^\Lam)$ and  $N^{\neq 0}_\Lam=N^{\neq 0}_\Lam(\ubx^\Lam )$
represent the number of empty and occupied sites in $\ubx^\Lam$. 
The transition probabilities $P_\Lam(\ubx^\Lam,\uby^\Lam )$ are non-zero only when the configurations 
$\ubx^\Lam$ and $\uby^\Lam$ differ at a single node $y$. Dropping the index $\Lam$,  
we have the following transitions and their probabilities:

\begin{enumerate} 
\item
If $y\in\Lam$ in $\ubx$ is occupied: $\ubx\to\ubx^{0,y}$ or $\ubx\to\ubx^{j,y}$,
\bea
P(\ubx,\ubx^{0,y})&=&\diy\frac{1-p}{(1-p+(p/q)Q(y,\ubx))\times |\Lam |},\\ 
P(\ubx,\ubx^{j,y})&=&\diy\frac{(p/q)\times{\mathbf 1}(\ubx^{j,y}\in\tA_\Lam (i))}{(1-p+(p/q)Q(y,\ubx ))\times |\Lam |}.
\eea

\item 
If $y\in\Lam$ in $\ubx$ is vacant: $\ubx\to \ubx^{j,y}$ or $\ubx\to \ubx$,
%
\bea\label{eq:5.4}
P(\ubx,\ubx^{j,y})&=&\diy \frac{(p/q)\times{\mathbf 1}(\ubx^{j,y}\in\tA_\Lam (i))}{(1-p+(p/q)Q(y,\ubx ))\times |\Lam |},\\
\label{eq:5.5}P(\ubx,\ubx)&=&\diy\frac{1-p}{(1-p+(p/q)Q(y,\ubx ))\times |\Lam |}.
\eea
\end{enumerate} 
Here $\ubx^{0,y}$ denotes the configuration where all but one sites in $\ubx$ preserve their status
and site $y$ becomes vacant. 
Next, $\ubx^{j,y}$ stands for the configuration where all sites in $\ubx$ preserve their status 
except for $y$ which becomes occupied with a particle of type $j$. Finally, $Q(y,\ubx)$ gives 
the number  of types $j$ for which ${\mathbf 1}(\ubx^{j,y}\in\tA_\Lam (i))=1$.

Using \eqref{eq:5.1}--\eqref{eq:5.5},  we get the detailed balance equation
\beq
\mu_\Lam (\ubx^\Lam ||i)P_\Lam (\ubx^\Lam ,\uby^\Lam )=\mu_\Lam (\uby^\Lam)P_\Lam(\uby^\Lam,\ubx^\Lam).
\eeq
This implies that the process iterations  lead to distribution $\mu_\Lam (\,\cdot\,||i)$, 
close to the PP $\mu (\,\cdot\,||i)$ for a large $\Lam$.

\begin{figure}[t]
\centering 
\includegraphics[width=9.5cm]{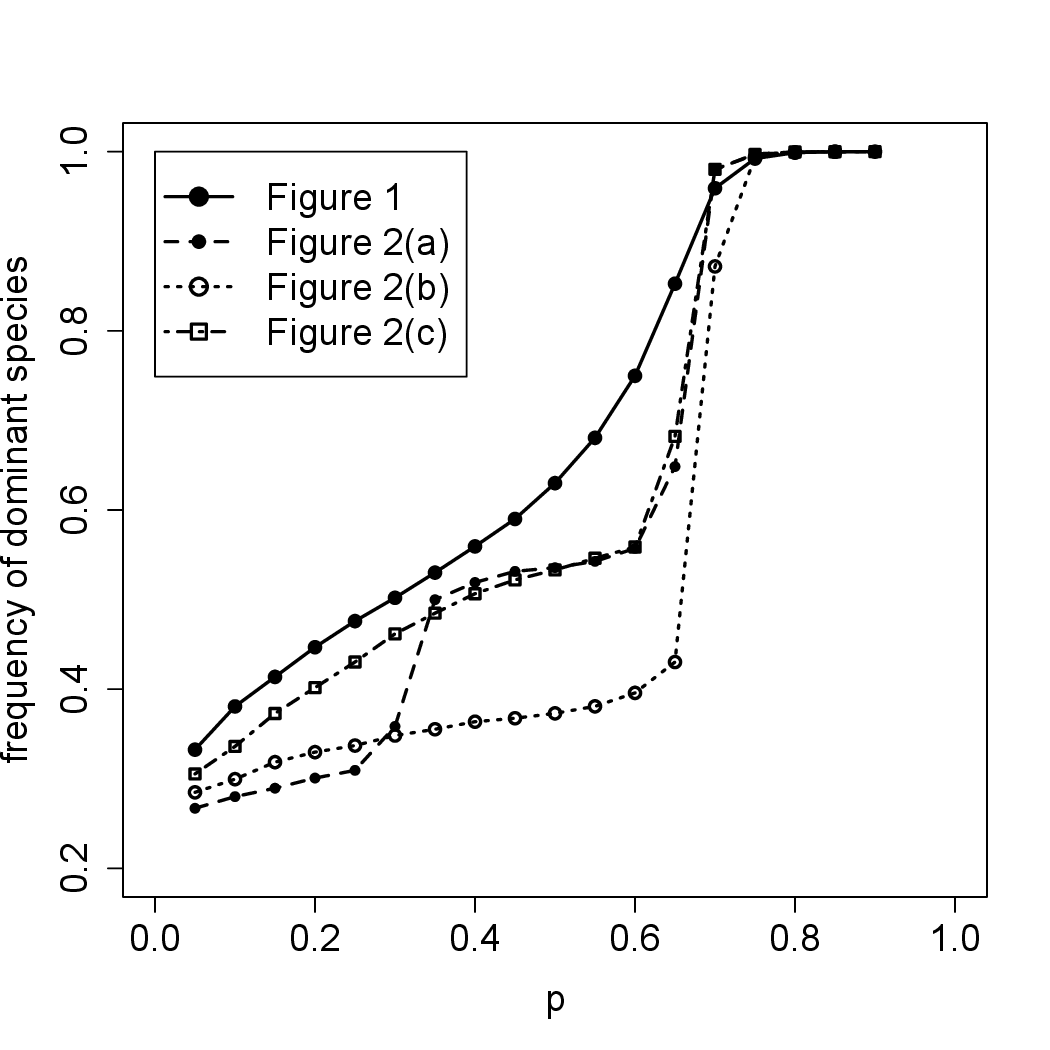}
\caption{Particle density for species $i$ in the measure $\mu (\,\cdot\,\|i)$ as a function of $p$. Shown are results for the configurations of Figure \ref{fig1} and Figure \ref{fig2} (a) -- (c).}
\label{fig3}
\end{figure}

Numerical simulations using the above spin-flip process were performed for $q=4$ species on a square  $200\times200$ in $\bbZ^2$ for various combinations of values of $p$ and collection $\mathcal{D}$. The number of iterations was $5 \times 10^8$. For a square $200\times200$ the equilibrium is generally reached at $\sim 10^7$ iterations. 

We have numerically verified the PPs for all examples shown in Figures \ref{fig1} and \ref{fig2}, where we choose 
$a(1)=2$ (green),  $a(2)=3$ (blue),  $a(3)=4$ (red),  $a(4)=5$ (yellow) and $p=0.8$. When there are 
several PPs, we fix the boundary conditions accordingly. In all cases we obtained the correct PP. 

As was said, for $p^*<p<1$ and $p\sim 0$ a PP has a different structure. This points at a \emph{critical phenomenon}  at some $p_{\rm{cr}}<p^*$. We record it numerically, with the relative frequency of the dominant species $i$ as a function of $p$. This is done for the configuration of Figure \ref{fig1} and Figure \ref{fig2} (a)--(c)  where we use the above values for EDs,  square size and iterations. The result are shown in Figure \ref{fig3}. One observes the value $p^*\approx 0.75$ for all cases after which the relative frequency of the dominant species becomes $\approx 1$. It is interesting to observe that while all four examples show a phase transition just below $p^*\approx 0.75$, the configuration of Figure \ref{fig2}(a) provides evidence for an intermediate phase with a second transition at a lower value of $p$. Note that in this case we have total symmetry between the four species and the result is qualitatively consistent with the case of equal EDs \cite{NL}. 

\begin{figure}[t]
\centering 
\includegraphics[width=9.5cm]{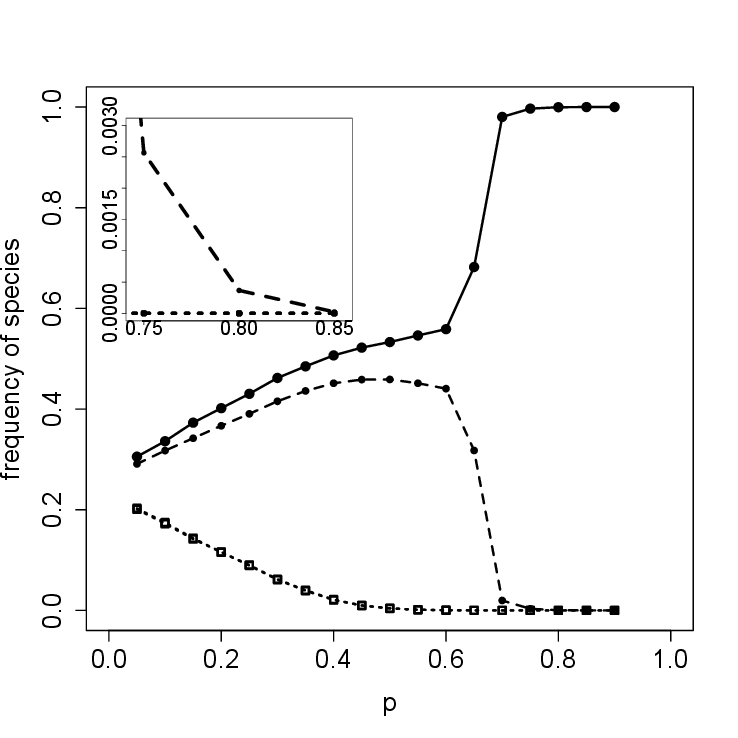}
\caption{Particle densities for four species as functions of $p$ for the picture in Figure \ref{fig2} (c). Shown are the dominant species $i=i_1$ (continuous) as well as $i_2$ (dashed) and $j_1,j_2$ (dotted, lying on top of each other). The inlay is a close up.}
\label{fig4}
\end{figure}

Another interesting aspect of the result of Proposition 1 is the relations between the densities of sub-dominant species. We analyse this numerically for the configuration of Figure \ref{fig2} (c) where we record the densities for all four species for varying $p$. The result is shown in Figure \ref{fig4}; it numerically verifies both the relations \eqref{eq:3.2A} and \eqref{eq:3.3} and confirms the inequalities \eqref{eq:4.11}.

\section{Discussion}
%
We have introduced a multi-type lattice WR-model with varying hard-core exclusion diameters. We identify the pure phases (PPs) in the case of four species for all collections of hard-core EDs. Our analysis shows that the PPs are dominated by the  ``most tolerant'' species, with smallest hard-core exclusion diameters (EDs). This is complemented with numerical results simulated using a spin-flip Markov process.

Regarding the analytical results we remark that it is in principle possible to go beyond four particle types or species. For $q\geq 5$, the particle types can again be placed at the vertices of a simplex in ${\bbR}^{q-1}$. As before, given a collection of hard-core EDs $\mathcal{D}$,  we can list them in an increasing order and define the set $\cS$ of ``lexicographically maximal'' 
particle types. For $p\sim 1$, any dominant type $i$ belongs to $\cS$. In particular, if  $\sharp\,\cS=1$,
i.e., $\cS$ is reduced to a singleton, there is a unique DLR measure. Also, if $\cS$ coincides 
with the whole of $\{1,\ldots ,q\}$,  every type $i$ will be dominant. But in general not every type $i\in\cS$ is dominant. An example is with five species and $7$ pair-wise distinct hard-core EDs, where $D(1,2) < D(1,3)=D(2,5) < D(1,4)=D(2,3) < D(1,5)=D(2,4) < D(4,5) < D(3,4) <
D(3,5)$. Here $\cS=\{1,2\}$, and type $1$ is dominant but  $2$ not; to determine this one needs to look at 
two-link paths along the edges of the five-vertex simplex (a pentagram). 

As to the simulations, it is interesting to view the spin-flip process as a dynamical model, viz., of spread of  cancer cells. Several lines in this direction are currently pursued. \\

{\emph{Acknowledgements --}}
The authors express their gratitude to Prof R. Chammas and Prof A. Ramos for providing references on cancer research and for numerous discussions on the applicability of the WR model in studies of tumor growth. This work is done under Grant 2011/51845-5 by FAPESP and Grant 2011.5.764.45.0 by USP. YS and IS thank IME, USP, for the hospitality. SZ was supported by FAPERJ (Grant 111.859/2012), CNPq (Grant 307700/2012-7) and PUC-Rio.


\section*{References}
 \begin{thebibliography}{10}


 \bibitem{WR}
B. Widom, J. S. Rowlison,
New model for the study of liquid-vapor phase transitions. 
{\it J. Chem. Phys.} {\bf 52} (1970) 1670.
 
 \bibitem{Ru}
D. Ruelle, Existence of a phase transition in a continuous classical system. 
{\it Phys. Rev. Lett.}, {\bf 27} (1971), 1040.
 
 \bibitem{LG}
J.L. Lebowitz, G. Gallavotti, 
 Phase Transitions in Binary Lattice Systems. 
{\it J. Math. Phys.} {\bf 12} (1971), 1129

 
\bibitem{RL} 
L.K. Runnels, J.L. Lebowitz,
Phase transitions of a multicomponent Widom--Rowlinson model.
 {\it J. Math. Phys.} {\bf 15} (1974) 1712.

\bibitem{CCK}
J.T. Chayes, L. Chayes, R. Kotecky.  {\it Comm. Math. Phys.}
{\bf 172} (1995), 449

\bibitem{LMNS}
J.L. Lebowitz, A. Mazel, P. Nielaba, L Samaj, 
 Ordering and Demixing Transitions in Multicomponent Widom-Rowlinson Models. 
{\it Phys. Rev. E} {\bf 52} (1995) 5985.

\bibitem{DS}
R. Dickman, G. Stellb,
Critical behavior of the Widom-Rowlinson lattice model. 
{\it J. Chem. Phys.}, {\bf 102} (1995), 8674.

\bibitem{NL}
P.~Nielaba, J.L.~Lebowitz, (1997),
Phase transitions in the multicomponent Widom-Rowlinson model and in hard cubes on the BCC-lattice. 
 arXiv:cond-mat/976305.


 \bibitem{GZ}
H.-O. Georgii, V. Zagrebnov,
 Entropy-driven phase transitions in multitype
lattice gas models. 
{\it J. Stat. Phys.}, {\bf 102} (2001), 35.



\bibitem{MSS}
A. Mazel, Y. Suhov, I. Stuhl, (2013),
A classical WR model with $q$ particle types.
submitted to {\it Com. Math. Phys.}
arXiv:1311.0020.

\bibitem{ID}
E. L. Rice, {\it Allelopathy}, (1984), Academic Press.

 
\bibitem{ZLXHH}
L. Zhang, Y.-K. Lau, W. Xia, G.N. Hortobagy, M.C. Hung,
Tyrosin kinase inhibitor emodin suppresses growth of HER-2/neu-overexpressing breast cancer cells
in athymic mice and sensitizes these cells to the inhibitory effect of paclitaxel.
{\it Clin. Cancer Res.} {\bf 5} (1999), 343.


\bibitem{App1} J. Hickson {\em et al}. 
Societal interactions in ovarian cancer metastasis: a quorum-sensing hypothesis. 
{\em Clin. Exp. Metastasis} {\bf 26} (2009) 67. 



\bibitem{Dobrushin}
R.L. Dobrushin. 
Description of a random field by means of conditional probabilities and the conditions governing its regularity. 
{\it Theor. Prob. Appl.} {\bf 13} (1968), 197.

\bibitem{LR}
O.E. Lanford, D. Ruelle. 
Observables at infinity and states with short range correlations in statistical mechanics. 
{\it Comm. Math. Phys.} {\bf 13} (1969), 194.
 

\end{thebibliography}
\end{document}